\documentclass[namedreferences]{SolarPhysics}
\usepackage{epsfig}
\usepackage{lscape}

\newcommand{\be}{\begin{equation}}
\newcommand{\ee}{\end{equation}}
\newcommand{\beq}{\begin{eqnarray}}
\newcommand{\eeq}{\end{eqnarray}}

\begin{document}
\begin{article}
\begin{opening}

\title{X-RAY POLARIZATION OF SOLAR FLARES MEASURED WITH RHESSI}

\author{E. \surname{SUAREZ-GARCIA}$^{1}$,
        W. \surname{HAJDAS}$^{1}$,
        C. \surname{WIGGER}$^{1}$,
        K. \surname{ARZNER}$^{1}$,
 	M. \surname{G\"UDEL}$^{1}$,        
        A. \surname{ZEHNDER}$^{1}$, and
	P. \surname{GRIGIS}$^{2}$}
	
\institute{$^1$Labor f\"ur Astrophysik, Paul Scherrer Institut, Villigen PSI, Switzerland \\
		$^2$Institute of Astronomy, Eidgen\"ossische Technische Hochschule, Zurich, Switzerland\\	
                  \email{estela.suarez@psi.ch}}

\runningauthor{SUAREZ-GARCIA ET AL.}
\runningtitle{X-RAY POLARIZATION OF SOLAR FLARES MEASURED WITH RHESSI}

\date{Received ; accepted }

\begin{abstract}

The degree of linear polarization in solar flares has not yet been precisely determined despite 
multiple attempts to measure it with different missions. The high energy range in particular has very rarely
been explored, due to its greater instrumental difficulties. We approached the subject using the Reuven Ramaty High Energy
Spectroscopic Imager (RHESSI) satellite to study 6 X-class and 1 M-class flares in the energy range between 100 keV and 350
keV. Using RHESSI as a polarimeter requires the application of strict cuts to the event list in order to extract those
photons that are Compton scattered between two detectors. Our measurements show polarization 
values between 2\% and 54\%, with errors ranging from 10\% to 26\% in $1\sigma$ level. In view
of the large uncertainties in both the magnitude and direction of the polarization vector, the results can only reject
source models with extreme properties.  

\end{abstract}

\end{opening}

\section{Introduction}
\label{Introduction} 

Measurements of the linear X-ray polarization in solar flares can provide essential information needed to
identify the processes responsible for the acceleration of particles and the emission of radiation. For photon
energies in the hard X-ray region, polarization is produced either through electron Brems\-strah\-lung or
Compton scattering. The degree of polarization is usually a complex function of the strength and topology of
the magnetic field. Thermal distributions of electron velocities result in small polarization values produced
mainly by the conduction-driven anisotropy of the electrons in the primary source \cite{emsliebrown80} with some contribution
from photons backscattered in the photosphere. Higher polarization levels are
expected from non-thermal anisotropic distributions of electrons that are accelerated in well ordered magnetic
fields. In addition, for such cases the observed polarization degree is related to the photon directivity (i.e.
the anisotropy of the emitted radiation), which depends both on the electron beaming details and on the viewing
angle. 

Starting in the late seventies, several non-thermal models of X-ray emission from solar flares were developed
(see \opencite{mcconnell02} and references therein). Generally the authors
assume a uniform magnetic field perpendicular to the solar surface and electrons being accelerated towards
the chromosphere (\opencite{elwerthaug70}; \opencite{haug72}; \opencite{brown72};
\opencite{bairamaty78}; \opencite{zharkova95}). These electrons, spiraling downwards along the magnetic
field lines, produce Brems\-strah\-lung radiation in collisions with hot plasma. The polarization in the
emitted X-rays is a function of the energy spectra of the electrons, their pitch angles (angle between their
velocity vector and the magnetic field), and the column
density distribution of the ambient plasma. Studying the spectral characteristics of the detected X-rays
provides  information about the electron energies, while the polarization is a very sensitive tool to sample
the electrons pitch angle distribution. Both for very small pitch angles, corresponding to high 
electron beaming, and for very large ones, the polarization degree can be equally high. The two cases can
be distinguished by their polarization direction: parallel and perpendicular to the magnetic
field line, respectively. The predicted polarization values can reach up to 60\% at energies above
50 keV, being even higher at low energies \cite{haug72}. After introducing more realistic pitch angle
distributions and taking into account photon backscattering processes in the photosphere, 
the expected polarization is reduced down to 20\% or 30\% (\opencite{brown72}; \opencite{bairamaty78}).

Several more complex non-thermal models have been developed as well. In one of them the magnetic field
structure was defined as a semicircular loop anchored in the chromosphere \cite{leachpetrosian83}.
This approach allowed studying the X-ray emissions separately from different parts of the loop. The highest
polarization could be produced at the top (up to 85\%), while the photons observed from the foot-points
(in the region of the dense chromosphere) would be polarized to the level of around 20\%.  

In general, lower energies are predicted to yield stronger polarization signals (\opencite{haug72};
\opencite{bairamaty78}; \opencite{leachpetrosian83}), although the strength of this relation can
vary depending on the model. In some recent theories an opposite trend has also been
reported \cite{zharkova95}. Experimental verification is usually difficult because the low-energy part
of the spectrum is strongly contaminated by a non-polarized thermal emission.

In all the cases the observed value of flare polarization is strongly dependent on the viewing angle. The
highest polarization values are expected for large angles of view, when the line of sight is perpendicular
to the magnetic field line. Thus, most theories predict higher polarization for flares located near
the solar limb. Similar behavior is also expected for the directivity of the flare emission: the
intensity of the emitted radiation should depend on the angle of view (\opencite{li94};
\opencite{li95}). Constraints related to the model assumptions favor two possible directions of the
polarization vector: either parallel to the plane defined by the magnetic field lines and the line of sight
or perpendicular to it (\opencite{bairamaty78}; \opencite{leachpetrosian83}; \opencite{zharkova95}).


Contrary to the intense theoretical work, only a few polarization measurements have been conducted in hard
X-rays. In this energy range, the commonly used technique is based on Compton scattering \cite{lei97}. 

First attempts were done at energies around 15 keV by \inlinecite{tindo70}, \inlinecite{tindo72}, and \inlinecite{tindo76}
using polarimeter instruments on board of several Intercosmos satellites. Although initial results \cite{tindo70}
showed a linear polarization of around 40\% $\pm$ 20\%, the later studies of different flares
(\opencite{tindo72}; \opencite{tindo76}) found their polarization degrees always compatible with zero. Moreover, the data suffered from limited
photon statistics and systematic errors related with the detector calibration. 

In measurements of solar flare polarization with the Reuven Ramaty High Energy Spectroscopic Imager (RHESSI), Compton
scattering can occur in a specially installed beryllium scatterer. This method can only be used at low energies (20--100
keV). Its details, together with RHESSI's polarimetric features, are described in \inlinecite{mcconnell02} and
\inlinecite{mcconnell04}.  

The latest measurements at energies below 100 keV have been performed  with the SPR-N instrument on board of
the Coronas-F satellite by \inlinecite{zhitnik06}. From a sample of 25 solar flares, these authors determined
upper limits on the polarization degree in the range from 8 to 40\% ($3\sigma$). Only for the single case of the
29 October 2003 flare they found a significant polarization degree, which increases from about 50\% at energies
$20-40~keV$, up to more than 70\% for the energy channel $60-100~keV$. 

Recently published, the only two measurements of solar flare polarization at high energies (0.2--1 MeV), show 1$\sigma$
values of 21\% $\pm$ 10\% and -11\% $\pm$ 5\% for one flare close to the limb and another near the Sun center,
respectively \cite{boggs06}. In their studies, the authors applied the polarimetric capabilities of the RHESSI satellite
in the coincidence mode, i.e. without using the Be-scatterer. 

In this paper, we present results in the energy range from 100 keV to 350 keV, obtained for seven
solar flares (X and M classes) also selected from the RHESSI instrument database. We used a method based on the
scattering of photons from detector to detector of RHESSI that has previously been applied for polarization studies of
gamma ray bursts (\opencite{coburnboggs03}; \opencite{wigger04}; \opencite{rutledgefox04}).

In \S \ref{sec:method} we explain how RHESSI can be used as a Compton polarimeter to study linear
X-ray polarization at energies $\geq$ 100 keV. The flares selected for analysis, and the criteria used for their
selections is detailed in \S \ref{sec:flselec}. Monte Carlo simulations were performed to calculate the
response of the instrument, and their results are discussed in \S \ref{sec:simulation}. The final polarization
results are described and compared with previous measurements in \S \ref{sec:results}. The interpretation of the
results is done in \S \ref{sec:interpretation} by comparing with theoretical predictions. Finally, a brief summary of
the conclusions of our work is given in \S \ref{sec:conclusions}.

\section{Method description}
\label{sec:method}

\subsection{Instrument}
\label{subsec:instrument}

The Reuven Ramaty High Energy Spectroscopic Imager \cite{lin02} was designed to observe solar flares from 3 keV
up to 17 MeV. It is capable of making spatially, 
spectrally and temporally resolved images of the Sun \cite{hurford02} using the rotation modulation principle
(\opencite{schnopper68}; \opencite{skinnerponman95}). The spacecraft is rotating with a period T $\approx$ 4 seconds.
The RHESSI angular position in the solar coordinate system is calculated using the satellite roll angle, which is
continuously monitored by the spacecraft aspect systems (\opencite{fivian02}; \opencite{hurfordcurtis02}).  

\begin{figure} 			
\centerline{\epsfig{file=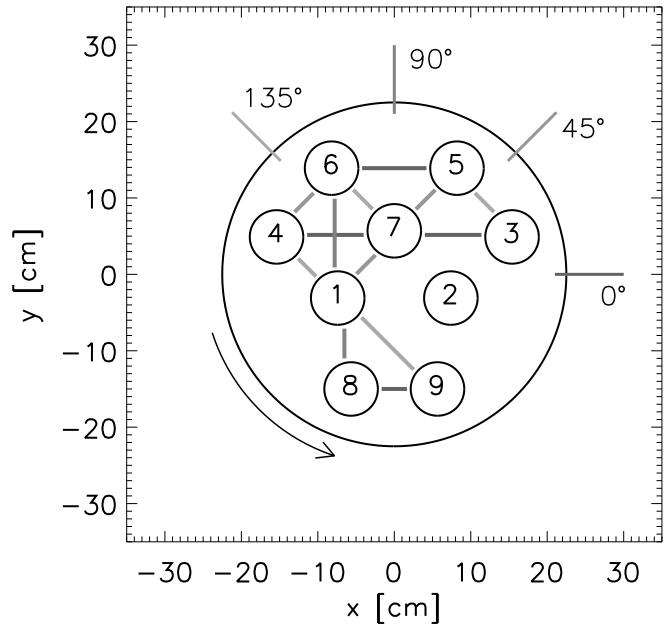,width=10cm,height=10cm}}
\caption{RHESSI detectors seen from the Sun. The \textit{grey lines} indicate the possible scattering directions in
	the RHESSI fixed coordinate system (\S \ref{subsec:polanalysis}).}
\label{fig:spectrometer}
\end{figure}

The RHESSI spectrometer \cite{smith02} consists of 9 cooled Germanium detectors which are split in
a thin front and a thick rear segment. Low-energy photons are mostly stopped in the front segments while high-energy
photons can pass through and reach the rear segments. The energy resolution is in the order of a few keV. The arrangement
of the 9 detectors in the spectrometer is sketched in Figure~\ref{fig:spectrometer}.

Each photon recorded by RHESSI is characterized by its arrival time, deposited energy, detector number and
segment. These parameters are stored in the RHESSI event list. The time
resolution of RHESSI is equal to one binary microsecond (1b$\mu$s = $2^{-20}$ sec). 

Polarization measurements are possible using photons that are Compton-scattered from one detector into another one, making
a signal in both of them. For such events, the effective area is very small, as most of the photons are either 
completely absorbed (in photopeak reactions) or scattered off the spectrometer \cite{hajdas05}. Therefore, only
a small percentage of all photons observed by RHESSI produces more than one entry in the event list. 
The fact that RHESSI rotates allows reducing the systematic errors in the polarization analysis. 

\subsection{Compton coincidences}
\label{subsec:coincidences}

The Compton scattering probability of photons on free electrons is given by the cross section \cite{kleinnishina29}:

\begin{equation}		
\frac{d\sigma}{d\Omega} = \frac{r_0^2}{2} \frac{E'^2}{E^2} \left(\frac{E}{E'} + \frac{E'}{E} - 
		2 \sin^2 (\theta) \cos^2 (\eta) \right),
\label{eq:klein-nishina}
\end{equation}

\noindent where $r_0$ is the classical electron radius, $\theta$ is the angle between photon infall and outfall
direction, and $\eta$ is the angle between the infall polarization and the outfall direction. $E$ and $E'$ are
the energies of the photon before and after the scattering, respectively. The cross section ($\sigma$)
is minimal if $\eta$ = 0$^\circ$ and maximal if $\eta$ = 90$^\circ$. The dependence of the cross section on $\eta$ is most
pronounced if $\theta$ = 90$^\circ$. As the RHESSI axis is pointing to the Sun, the detectors plane is perpendicular to
the photon infall direction, and therefore $\theta \approx 90^\circ$ for detector-to-detector scattering.  

In order to extract Compton scattered photons from the RHESSI event list, several cuts were applied (see
Table~\ref{tb:cuts}).

\begin{table}[ht] 		
\begin{tabular}{ll} 
Criteria       			      	& Values\\\hline
Time interval  		      		& Between 1 min. and 4 min. around the flare peak\\
Coincidence width 		 	& 1 b$\mu$s \\
Single event energy 	      		& 25 keV $< E_i <$ 300 keV \\
Coincidence energy (sum)		& 100 keV $< E_i + E_j <$ 350 keV \\
Kinematical cut     	      		& $\theta = 90^\circ \pm 45^\circ $\\
Detector segments 	      		& Rear segments (without detector 2) \\
Pairs selection 	      		& Neighbor detectors \\
Multiplicity		      		& Two-event coincidences only \\
\caption{Selection of cuts introduced to extract Compton scattered photons from the event list.} 
\label{tb:cuts}
\end{tabular}
\end{table} 

As most of the high energy photons pass through the front segments without any interaction, only the rear
segments were used for the polarization analysis. This condition ensures that Compton scattering
events happen within the same b$\mu$s (\inlinecite{wigger04}, Equation (8)). Detector 2 was not taken into account as it
operates in a different way than the others \cite{smith02}.

When more than two events happen at the same time, information about the polarization
direction is lost. Therefore we selected photons which were registered in exactly two detectors, calling them
'coincidences'. Furthermore, only coincidences between neighboring detectors were chosen, due to the very low
probability of direct scattering between remote detectors (see Figure~\ref{fig:spectrometer}).

The sum of the two energies of the coincidence ($E_i$ and $E_j$) had to fall into our selected range between 
100 and 350 keV. In addition, the energy cut imposed on the individual detectors was in the range 25--300 keV. Below
this region, the detector noise and the level of accidental coincidences (see~\S\ref{subsec:bg}) increased very
strongly. A final constraint was applied by introducing the so-called kinematical cut which excludes 
coincidences that are incompatible with Compton scattering, using the kinematical relation between the
scattering angle $\theta$ and the observed energies. The kinematical cut was especially useful to reject
photons scattered from the Earth atmosphere (see \S \ref{sec:simulation}).  
  
Applying the above cuts, we obtain a raw list of coincidences which is still contaminated by background events of
different origin.

\subsection{Background subtraction}
\label{subsec:bg}

Two major sources of background were taken into account in the following analysis. The first are accidental
coincidences which occur when two independent solar photons are simultaneously
detected. The second kind of background is not related to the flare itself but produced by cosmic rays and the
cosmic gamma ray background.

The rate of accidental coincidences is proportional to the square of the incoming photon flux. In order to obtain the
number of accidental coincidences ($N_{\rm acc}$) we repeated the same procedure that was used to find the raw list of
coincidences (Table~\ref{tb:cuts}), but taking pairs of events which are time delayed by 20
b$\mu$s $< dt <$ 30 b$\mu$s (see \opencite{wigger04}). As the typical detector dead time is in the order of several
microseconds, the delay length was chosen to be longer than this. The rate of accidental coincidences over this range is
approximately constant and was used for background subtraction.   

\begin{figure}			
\centerline{\epsfig{file=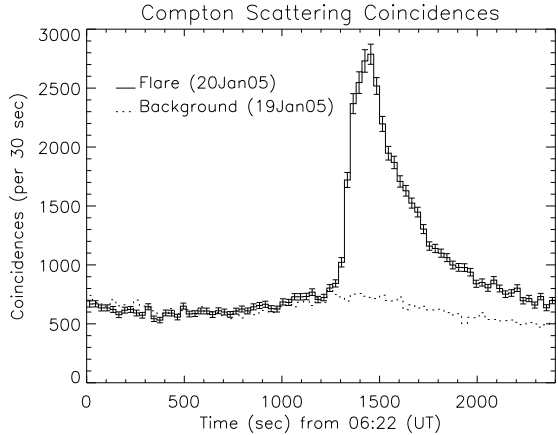,width=8cm,height=6cm}}
\caption{Signal and background coincidences vs. time for the flare on 20 January 2005 together with
	the background levels measured approximately one day before at the same position.}
\label{fig:bglevel}
\end{figure}

The second kind of background was determined either before or after the flare. For this purpose we selected time
intervals shifted by 18 orbits, i.e. 24 hours, before or after the flare peak. In this way, the spacecraft
geomagnetic coordinates were similar (see first part of Table~\ref{tb:geocor_Ncoinc} for an example), and
the systematic effects coming from the background variations along the orbit were strongly reduced. Proper
subtraction of this background requires some additional conditions: no other flare or high energy event should be
present in that period, and the spacecraft operational status regarding attenuator state and decimation logic should
be close to the one in the moment of the flare observation. The total number of background coincidences obtained in
this way is denoted by $N_{bg,tot}$. The period selected for background subtraction contains a (usually
negligible) number $N_{bg,acc}$ of accidental coincidences. It has to be subtracted from $N_{bg,tot}$ because it
is already included in $N_{acc}$ calculated during the flare peak. The number of coincidences produced by the
non-flare related background is then $N_{bg} = N_{bg,tot} - N_{bg,acc}$. As an example, the time evolution of the
background signal observed one day before the 20 January 2005 flare is displayed together with the flare in
Figure~\ref{fig:bglevel}. It  reproduces very well the background levels observed before and after the flare peak.

\begin{table}[ht] 		
\begin{tabular}{ccc}
				& PEAK OF FLARE 	& BACKGROUND\\
Starting time			& 06:43:00 20-Jan-2005	& 06:42:59 19-Jan-2005\\\hline
RA($^\circ$)			& -76.07 		& -80.26\\
Dec($^\circ$)			& 37.91 		& 37.91\\\hline
Total coincidences		& $43313 \pm 208$	& $5874 \pm 77$\\
Accidental coincidences		& $26907 \pm 35$        & $66 \pm 2$ \\
Solar flare Compton scattering coincidences         & \hspace{2cm} 10598 $\pm$ 225 \hspace{-1cm} \\\hline
\end{tabular}
\caption{Geomagnetic coordinates of RHESSI satellite and numbers of coincidences determined at the peak
	of the flare on 20 January 2005 and during the background measurement period. The data
	collecting time was 240 seconds.}  
\label{tb:geocor_Ncoinc}
\end{table}

 Finally, the rate of Compton Scattering events was calculated for any time $t$ according to   

\begin{equation} 		
  N_{C}(t) = N_{tot}(t) - N_{acc}(t) - N_{bg}(t + \Gamma),
\label{eq:nC}
\end{equation}

\noindent where $\Gamma$ is chosen such that both the geomagnetic coordinates and the orientation (roll
angle) of RHESSI are in the background period as close as possible to the ones during the flare peak. $N_{tot}$ is the
number of coincidences in the flare peak according to Table~\ref{tb:cuts}, $N_{bg}$ is the corresponding number of
background events from flare-unrelated photons, and $N_{acc}$ is the number of accidental coincidences.

\subsection{Polarization analysis}
\label{subsec:polanalysis}

For each coincidence, the line connecting the two detectors involved in the Compton scattering process
defines the angle used to construct the modulation curve. According to the RHESSI fixed coordinate system, there are
four possible scattering directions $d$ (labeled $0^\circ, 45^\circ, 90^\circ$, and $135^\circ$) between neighboring
detectors (Figure~\ref{fig:spectrometer}). In each direction, the number $N_d(\tau_i)$ of coincidences 
per time bin $\tau_i$ (with $\tau_i = t_i$ MOD $T$) was divided by their sum in the
same direction ($n_d(\tau_i) = \frac{N_d(\tau_i)}{\sum_{i}{N_d(\tau_i)}}$). Such normalization is needed in order
to eliminate the dependence on the number of detector pairs which is different for each direction (see
Figure~\ref{fig:spectrometer}). 

The RHESSI roll period T was measured using the spacecraft roll aspect system (\opencite{fivian02};
\opencite{hurfordcurtis02}) which provides the RHESSI angular position with an accuracy of 1 arcmin. We obtained the satellite
rotation period for each flare using a linear fit. In order to verify its stability and check for possible drifts, T
was calculated in 1 sec steps over a 720 sec interval containing the flare. The maximum variations between the
measurements were below 1 ms, thus negligible comparing with the bin size used in the polarization analysis ($\approx$ 167 ms).
It was also found that the drift of the period is smaller than $(2.1 \pm 2.0)\cdot 10^{-7}$. It corresponds to a change in the
rotation period by less than 0.05 ms, giving an upper limit of $0.53^\circ$ (3$\sigma$) on the possible phase shift for
the longest time interval used in the analysis. 

The asymmetry curve can be constructed as follows:

\begin{equation} 		
  A_{0-90}(\tau_i) = \frac{n_0(\tau_i)-n_{90}(\tau_i)}{n_{0}(\tau_i)+n_{90}(\tau_i)}.
\label{eq:asym}
\end{equation}
 
\noindent The coincidences occurred in the directions $45^\circ$ and $135^\circ$ were properly shifted and
included in the equation above. Using asymmetry to determine the polarization minimizes the effect of the lightcurve 
variations and grouping to have only two directions improves the statistics.   

In order to relate the time variable $\tau$ with angular directions in the Sun, we used the relation:
$\eta_i = -\frac{2\pi}{T} \cdot \tau_i + \eta_0$. By convention, $\eta = 0$ corresponds to solar West, and $\eta = \pi/2$
corresponds to the solar North. $\eta_0$ is the angular position of RHESSI X-axis ($0^\circ$ direction from
Figure~\ref{fig:spectrometer}) with respect to the solar West in the moment when measurement started. Trough this
coordinate transformation, we obtained $A_{0-90}(\tau)$ in heliocentric coordinates: $A(\eta)$.

Due to the sinusoidal dependence of the Compton cross section for the scattering of a polarized photon (Equation
(\ref{eq:klein-nishina})), the asymmetry curve is also a sinusoidal function with period equal to a half of the RHESSI 
rotation. It can be represented by a function:

\begin{equation} 		
   A(\eta) = \mu_{p} \cdot \mbox{cos}\left(2(\eta - \phi + \pi/2)\right),
\label{eq:fitfun}
\end{equation}

\noindent where the amplitude $\mu_{p}$ is a positively defined value equal to the flare modulation factor. The
phase $\phi$ is the polarization angle from the flare. 

 Comparing the experimental amplitude $\mu_{p}$ with the modulation factor $\mu_{100}$ from
Monte Carlo simulations for a 100\% polarized flux (\S\ref{sec:simulation}), allows to determine the
polarization degree of the solar flare ($\Pi$) \cite{mcconnell02}:  

\begin{equation}  		
  \Pi = \frac{\mu_{p}}{\mu_{100}}.
\label{eq:Poldeg}
\end{equation}

Two parameters, $\Pi$ and $\phi$, fully describe the polarization state of a solar flare and are needed
for comparison with theoretical predictions.  

\section{Flares selection}
\label{sec:flselec}

\begin{figure}	 		
\centerline{\epsfig{file=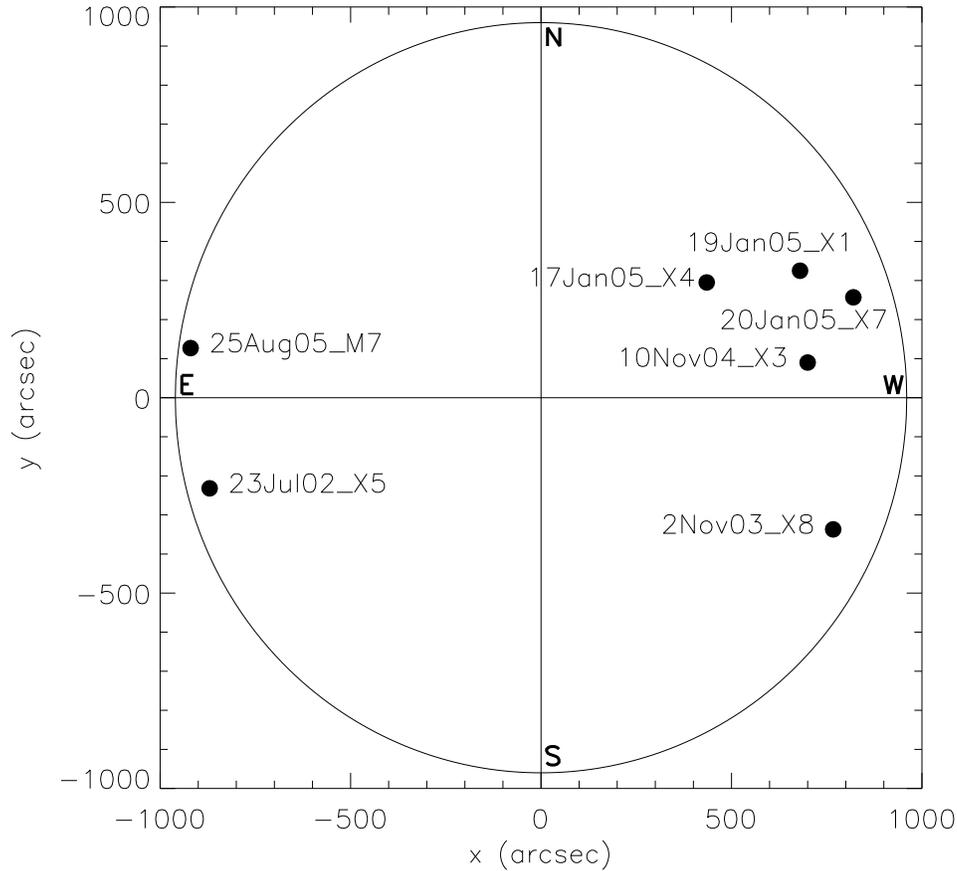,width=13cm,height=12.5cm}}
\caption{Distribution of the flares in the solar disc. Their position was extracted from images taken with RHESSI.} 
\label{fig:mapsun_dots}
\end{figure}

The following criteria were applied to select the flares for polarization analysis: large intensity,
strong high energy component and negligible contamination with particles (either from the flare 
itself or from the radiation belts). Also, since theory predicts highest polarization close to the solar limb
(\opencite{brown72}; \opencite{bairamaty78}; \opencite{leachpetrosian83}), we focused on limb-close flares. After
applying all these conditions, we were able to select six X and one M class flares, five of them located within
less than 120 arcsec from the solar limb (Figure~\ref{fig:mapsun_dots}). 

\begin{figure}			
\centerline{\epsfig{file=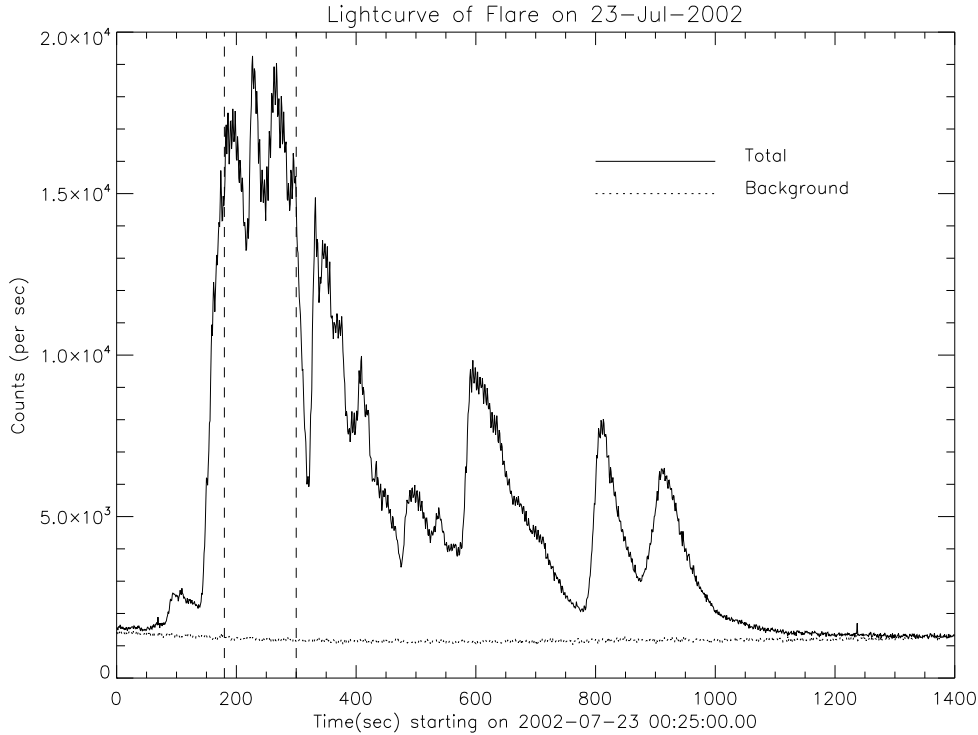,width=13cm,height=10cm}}
\caption{Ligthcurve of the solar flare on 23 July 2002 and background selected for it. The number of single
	events registered in rear segments of all detectors (except number 2), with energies between
	25 keV and 300 keV is plotted as a function of time. The part of the flare used for the polarization
	measurement is enclosed by the two \textit{vertical dashed lines}.} 
\label{fg:lcall}
\end{figure}

In order to study polarization in the most explosive part of the energy release, only the peak of the flare was
chosen (see Figure~\ref{fg:lcall}). The time period for analysis  varied between one and four minutes
depending on the duration of the flare peak.

\begin{figure}			
\centerline{\epsfig{file=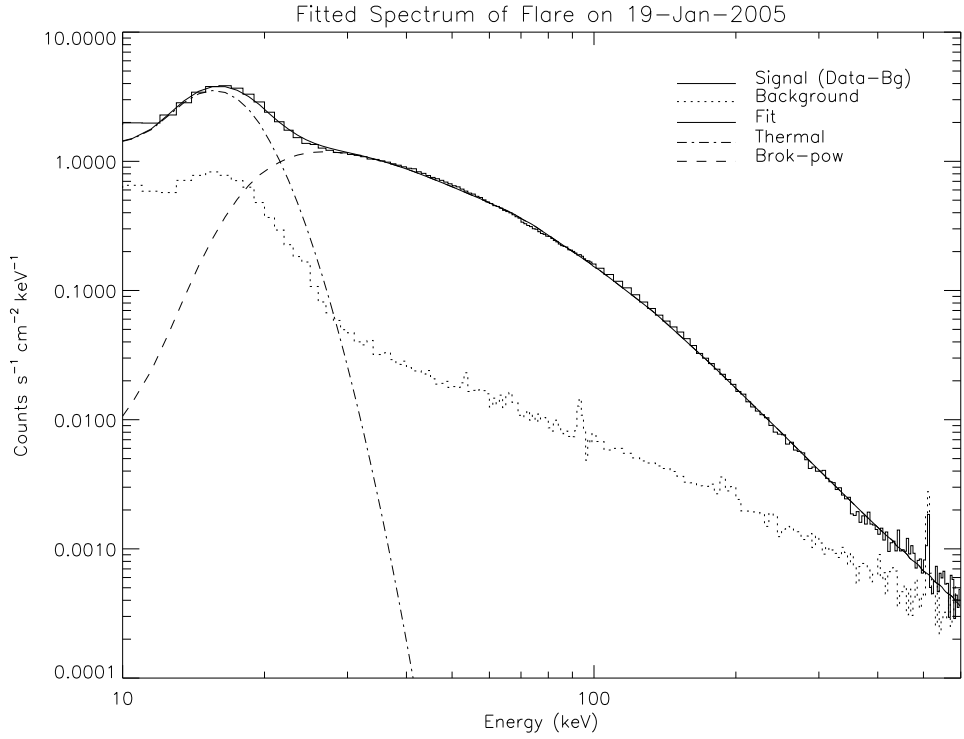,width=13cm,height=10cm}}
\caption{Background-subtracted spectrum of the flare on 19 January 2005. The fit was made for
	energies between 12 keV and 600 keV combining a thermal curve and a broken power law. Thermal
	component is negligible above 40 keV, and the background becomes dominant around 500 keV.}
\label{fig:fitspec}
\end{figure}

The spectrum of each flare was analyzed to find the energy ranges of different emission mechanisms. For
this purpose, a fit was performed, using the RHESSI OSPEX fitting tool \cite{tolbert06}, with 
a combination of a thermal Bremsstrahlung curve and a broken power law (Figure~\ref{fig:fitspec}). In all cases,
the thermal contribution was found to be negligible at energies above 50 keV. For further analysis we selected
the non-thermal Bremsstrahlung region (100--350 keV). The spectral indices are given in Table~\ref{tb:flares} and
correspond to a single power law fit of that part of the flare. These values were subsequently used in the
simulations performed to determine the instrumental response function and its polarization modulation factor
(see~\S\ref{sec:simulation}). Although photons with energies between 50 and 100 keV are already in the
non-thermal emission region, their interaction in the RHESSI detectors is governed by photoelectric
absorption, leaving only a marginal number of Compton scattering events. On the high-energy side, the 
threshold value was chosen to avoid regions dominated by background.

The main parameters that describe the flares selected are summarized in Table~\ref{tb:flares}.

\begin{landscape}	
\begin{table}[p] 		
\begin{tabular}{llllllll}
\hline
Flare number (RHESSI)	& 2072301	& 3110221	& 4111002	& 5011710	& 5011911	& 5012005	& 5082502	\\
Date			& 23 Jul. 2002	& 2 Nov. 2003	& 10 Nov. 2004	& 17 Jan. 2005	& 19 Jan. 2005	& 20 Jan. 2005	& 25 Aug. 2005 	\\
Start time 		& 00:18:00	& 17:03:00	& 01:59:00	& 09:35:36	& 07:57:20	& 06:21:24	& 04:33:48	\\
End time		& 01:15:44	& 18:00:36	& 02:36:52	& 10:38:48	& 09:03:32	& 07:27:04	& 04:55:56	\\
Duration (s)		& 3468 		& 3456		& 2272		& 3792		& 3972		& 3940	 	& 1328		\\
Class			& X4.8		& X8.4		& X2.6		& X4.0 		& X1.4		& X7.1		& M7.0		\\
(x, y) (arcsec) 		& (-875, -226) 	& (770, -320)	& (716, 99)	& (424, 312)	& (689, 325)	& (833, 245) 	& (-936, 120)	\\
Radial distance (arcsec) & 904	& 834		& 723		& 526		& 730		& 868	 	& 944		\\\hline
Start analysis		& 00:28:00 	& 17:15:40	& 02:08:30	& 09:43:20	& 08:24:40	& 06:43:00  	& 04:36:00	\\
Analyzed interval      	& 120		& 240		& 150		& 60		& 150		& 240		& 240		\\
Counts (s$^{-1}$) (E $>$ 25 keV)	& 67421 & 126440	& 28296		& 50001		& 20798		& 128520	& 21804		\\
Spectral index (100--350 keV)	& 3.1 	& 3.5		& 3.4		& 3.8		& 2.9		&3.2 		& 3.2		\\\hline

\end{tabular}
\caption{Description of the main characteristics of the flares studied.} 
\label{tb:flares}
\end{table} 
\end{landscape}	


\section{Simulations}
\label{sec:simulation}

\begin{figure} 			
\centerline{\epsfig{file=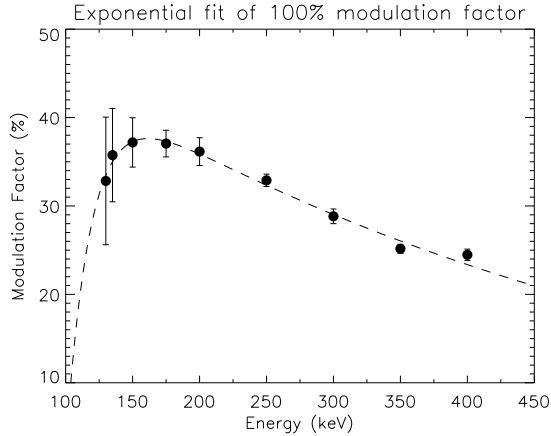,width=8cm,height=6cm}}
\caption{Simulated energy dependence of $\mu_{100}$. The \textit{dashed line} corresponds to the fit function: $\mu_{100}=56\cdot
	e^{-E/460} - 8668\cdot e^{-E/19}$. }
\label{fig:mu100fit}
\end{figure}

Monte Carlo simulations have been performed to calculate the response of the RHESSI polarimeter to a 
100\% polarized solar flare. For this purpose, the exact mass model of the whole
satellite and its germanium spectrometer has been constructed and implemented in the $GEANT$ 3.21 simulation
code \cite{cern94}. The spacecraft was illuminated by a uniform beam of photons coming parallel
to the RHESSI rotation axis. Each incoming photon was fully tracked and all energy depositions made on its
path in any of RHESSI detectors were recorded. In this way a simulated event list was created and subsequently
used to determine the modulation factors. The procedure to extract the
modulation factors from the event list was the same as for the analysis of the solar flares (\S\ref{sec:method}), but
neglecting the accidental coincidences. We made two
sets of simulations:  

\begin{enumerate}

\item The modulation factors for 100\% polarized emissions were determined for each flare using a photon 
energy distribution with spectral indices as given in Table~\ref{tb:flares}. The
incoming photons were 100\% polarized and their  
energy range was either 100--600 keV or 80--350 keV. It was found that the contribution from
photons above 350 keV to the modulation curve in our energy range (100--350 keV) was negligible.
An example of a modulation curve is shown in Figure~\ref{fig:asysim}.  

\begin{figure} 			
\centerline{\epsfig{file=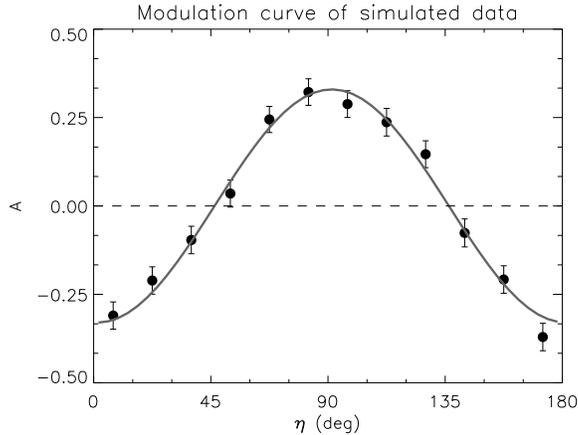,width=8cm,height=6cm}}
\caption{Simulated asymmetry plot for a 100\% polarized flare with spectral index -3.1.}
\label{fig:asysim}
\end{figure}

\item In the second case we took monoenergetic photons with $100\%$ polarization at thirteen energies between 100
keV and 1000 keV (Figure~\ref{fig:mu100fit}). The mean modulation factor of each flare was
subsequently calculated by averaging the monoenergetic modulation factors. The weights used for the average were
proportional to the number of coincidences per energy bin. $\mu_{100}$ is maximum around
170 keV. It decreases for lower energies due to the low energy threshold (25 keV) which only allows detecting recoil
electrons from photons scattered at very large angles. At such angles, modulation factor of the instrument is very
small, in accordance to Equation~(\ref{eq:klein-nishina}). Above 170 keV $\mu_{100}$ diminishes with energy following
the polarization sensitivity based on Equation~(\ref{eq:klein-nishina}) for photons scattered around $90^\circ$.
Using two exponential functions with properly chosen coefficients reproduces the above features in a simple way
(see fit function in the capture of Figure~\ref{fig:mu100fit}).  
\end{enumerate}

Both approaches gave the same results. For example, for the case of the 23 July 2002 flare, with an
spectral index equal to -3.1, we obtained with the first method $\mu_{100}=32.8\pm1.6\%$ and with the
second one $\mu_{100}=32.4\pm5.4\%$.

The analysis of the solar flare lightcurves from single detectors revealed periodic structures that can be
attributed to photons reaching RHESSI after being scattered by the Earth's atmosphere. The
contamination of the modulation curves by such photons was computed with another set of Monte Carlo
simulations. For this purpose a simplified system consisting of the satellite and the Earth with its
atmosphere was constructed. The Earth was represented by a solid sphere with twelve layers of atmosphere
extending up to about 50 km above the surface. The mass as well as the chemical composition of all the atmospheric
sheets were equal, while the density varied in accordance with their height.

\begin{figure} 			
\centerline{\epsfig{file=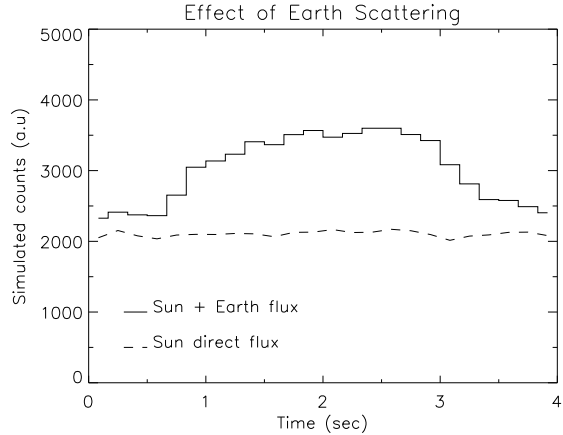,width=8cm,height=6cm}}
\caption{Simulations of scattering by the Earth's atmosphere. The figure shows single events with energies between
	100 keV and 350 keV arriving at the rear part of detector number 4. The \textit{solid line} shows the total photon
	flux, while the flux coming directly from the Sun appears as a \textit{dashed line}.} 
\label{fig:earthscatt}
\end{figure}

Simulations of Earth scattering were performed using unpolarized photons with spectral indices typical
for the analyzed flares and their corresponding angular positions between RHESSI, the Earth and the Sun.
The largest fraction of  photons detected from the atmosphere was found when the Earth reached the angle
of $90^\circ$ with respect to the RHESSI-Sun direction. In the energy range 100 keV-350 keV, up to 30\%
of the observed photons were coming from the Earth, producing a strong modulation in the single-event
lightcurves (Figure~\ref{fig:earthscatt}, solid line). The influence of such photons on the asymmetry curves,
extracted from coincidences, was much smaller. Most of the Earth scattering caused only accidental
coincidences or did not pass the kinematical cut. Finally, the contamination by the Earth-scattered photons was, in the
worst case, less than 8\%. Considering their low level and flat distribution along the modulation curves, the modifications of the measured modulation 
factors were negligible compared to the overall statistical error.

The presence of the grids above the detectors could produce periodic structures also in the coincidence lightcurves and
therefore this effect was carefully studied with Monte Carlo simulations. Those sources which are not in the center of
the Sun are, during RHESSI rotation, intermittently obscured by the grids. This produces in their single-event
lightcurves a modulated profile with $180^\circ$ periodicity \cite{hurford02} and such a pattern could be mixed up with
the real polarization signal. Firstly, the lightcurve modulation is strongly reduced for high energy photons reaching the
rear detectors. Secondly, the width of the angular bin used for polarization analysis is large enough to average all
count-rate variations caused by the grid modulation. The lightcurve variations further cancel out since the grids of
detector pairs used for coincidences are generally aligned in different directions.  

We performed simulations of a flare situated in the solar limb where the grid effect would be the strongest. Fine grids
have been approximated keeping the slit/slat width ratio while the coarse ones were exactly implemented. The modulation
factor obtained for a non-polarized flux of $10^8$ photons in the 100-350 keV energy range and with a power law spectrum
index of -3.1, was $2.5\%\pm1.9\%$. This is similar to the signal measured from a non-polarized source situated in the
center of the sun, where grids do not cause any modulation. Therefore, the grids effect can be neglected.    

Notice that due to the positive definiteness of the polarization degree, even an unpolarized signal
gives a non-vanishing amplitude in the asymmetry plot when applying to our analysis. From a 0\% polarized simulated
flare with around 8500 coincidences, we obtained a modulation factor equal to $(3.4\pm1.6)\%$.


\section{Results}
\label{sec:results}

\subsection{RHESSI polarization measurements}
\label{subsec:RHESSI}

\begin{figure}			
\hbox{\epsfig{file=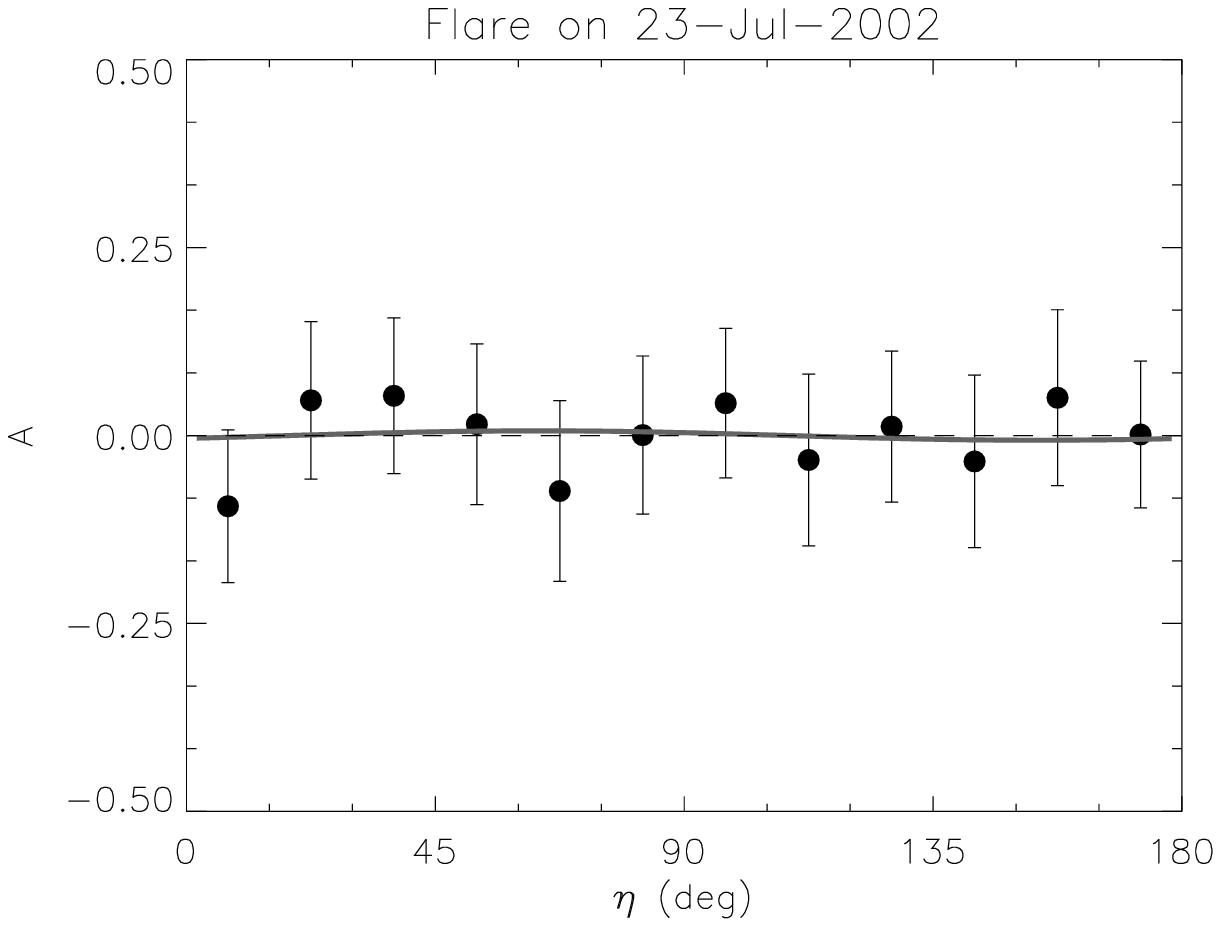,width=6.5cm}
	\epsfig{file=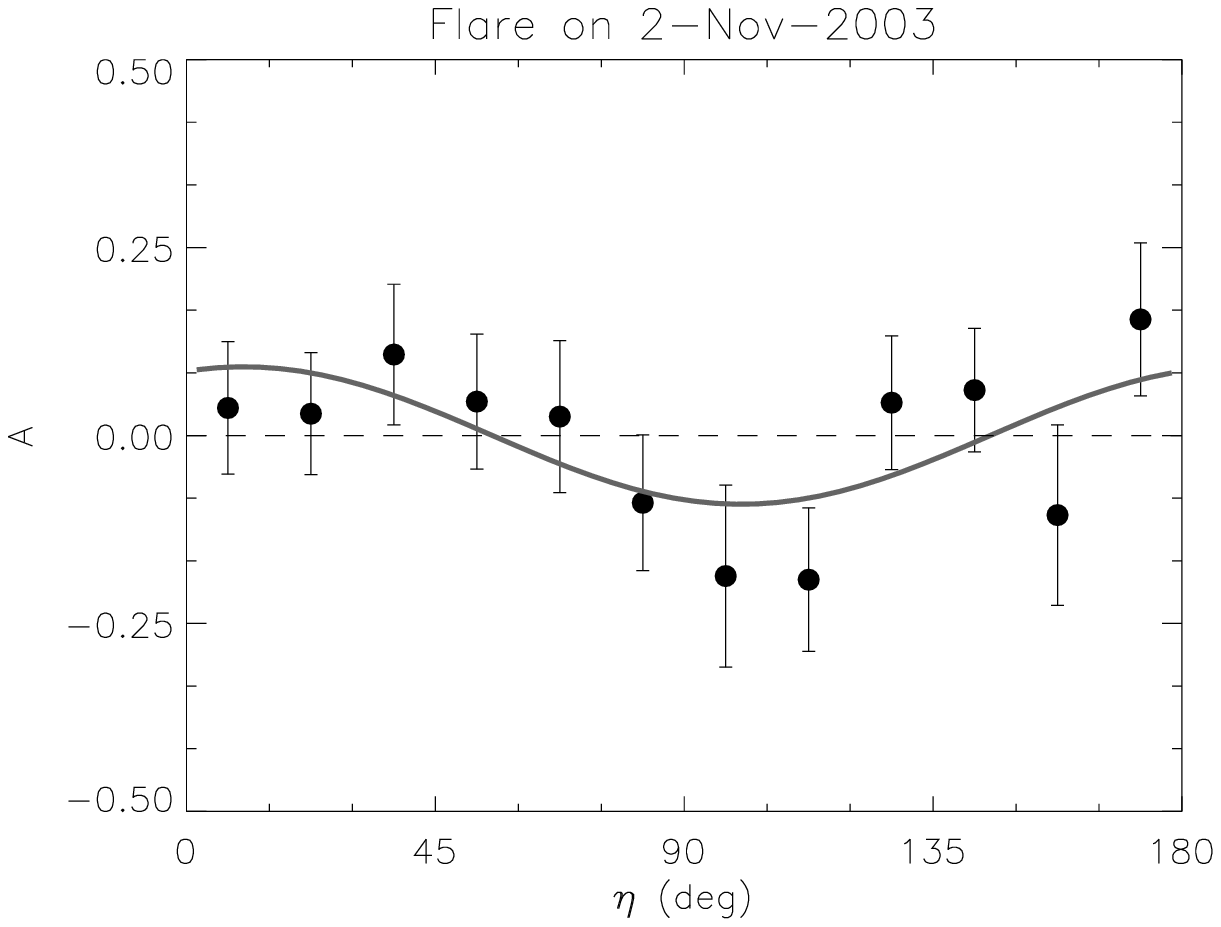,width=6.5cm}}
\hbox{\epsfig{file=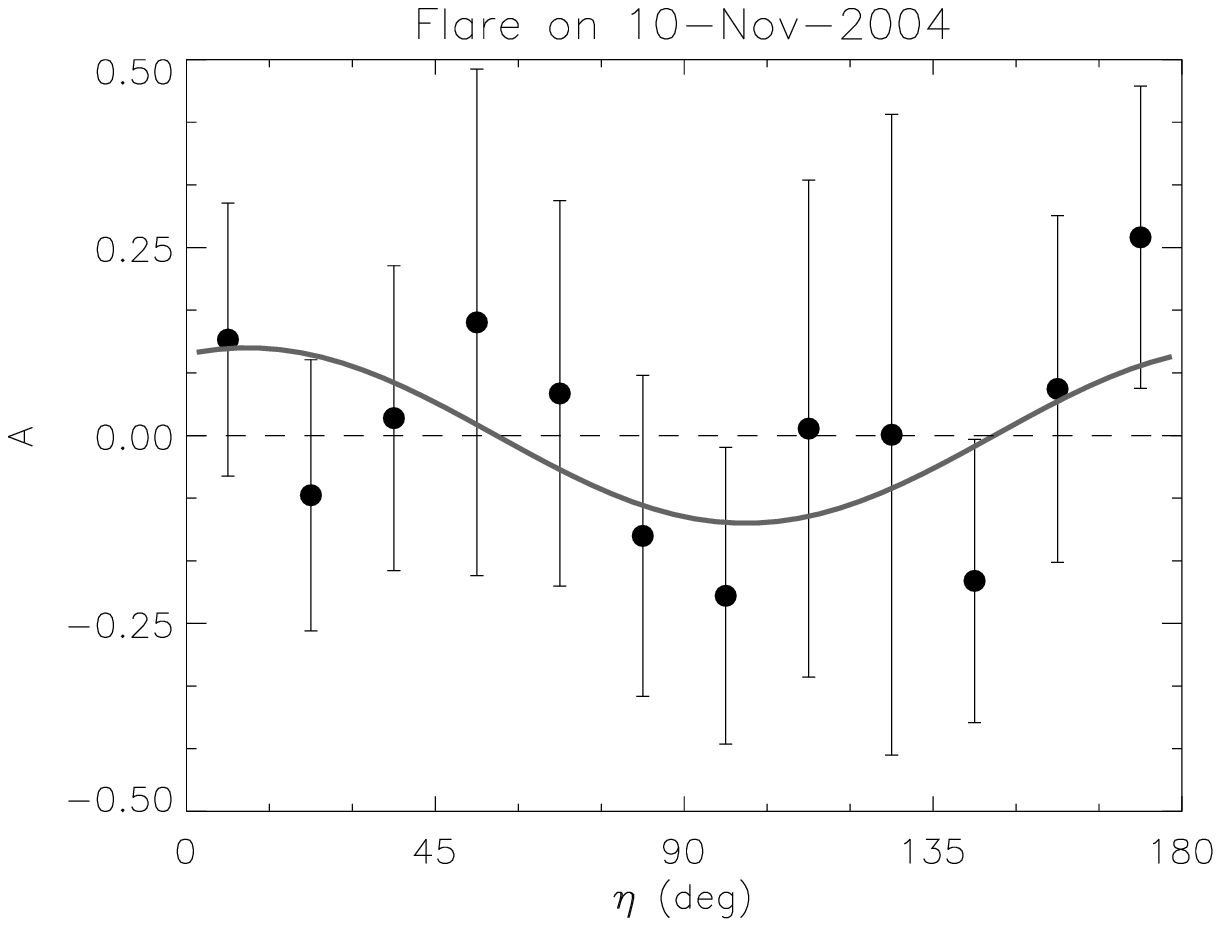,width=6.5cm}
	\epsfig{file=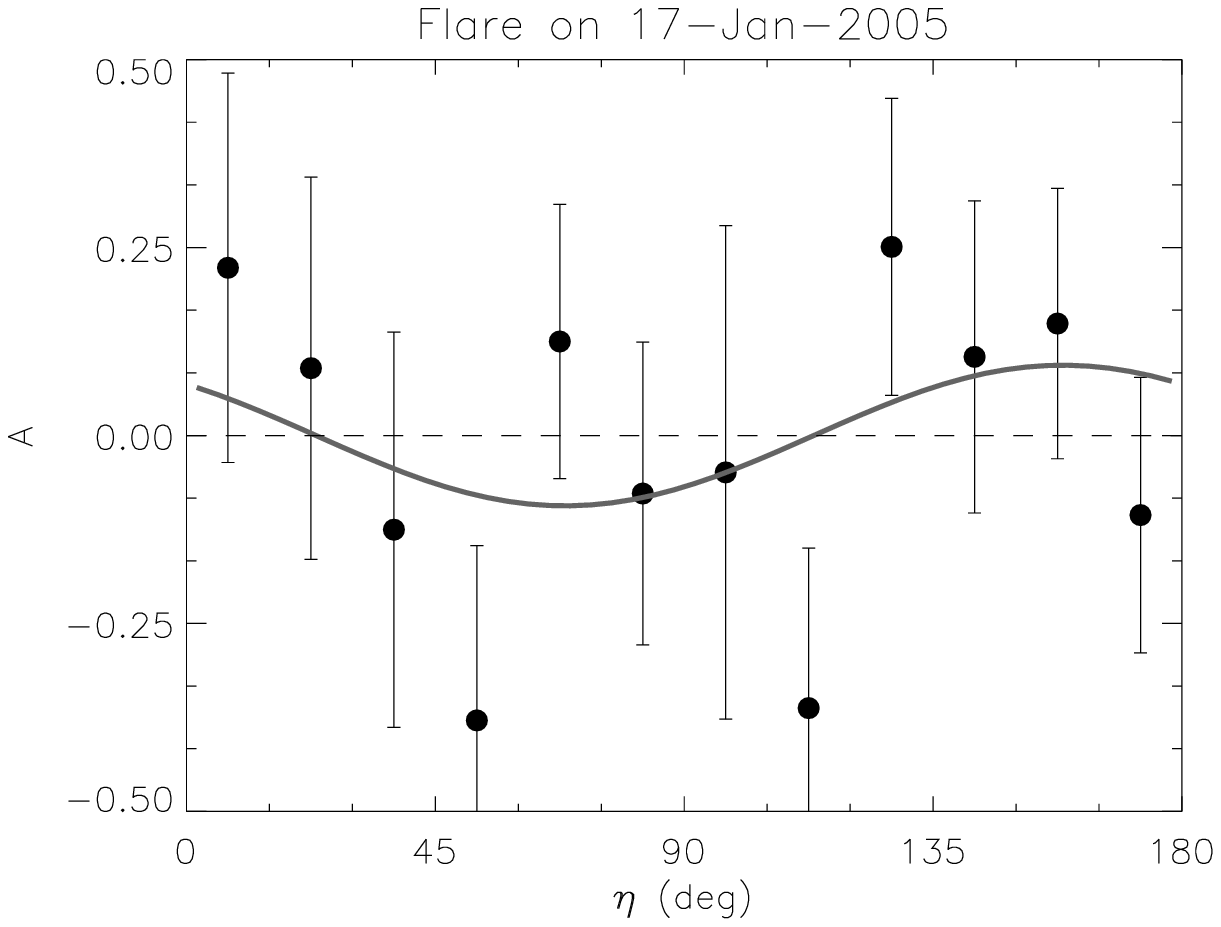,width=6.5cm}}
\hbox{\epsfig{file=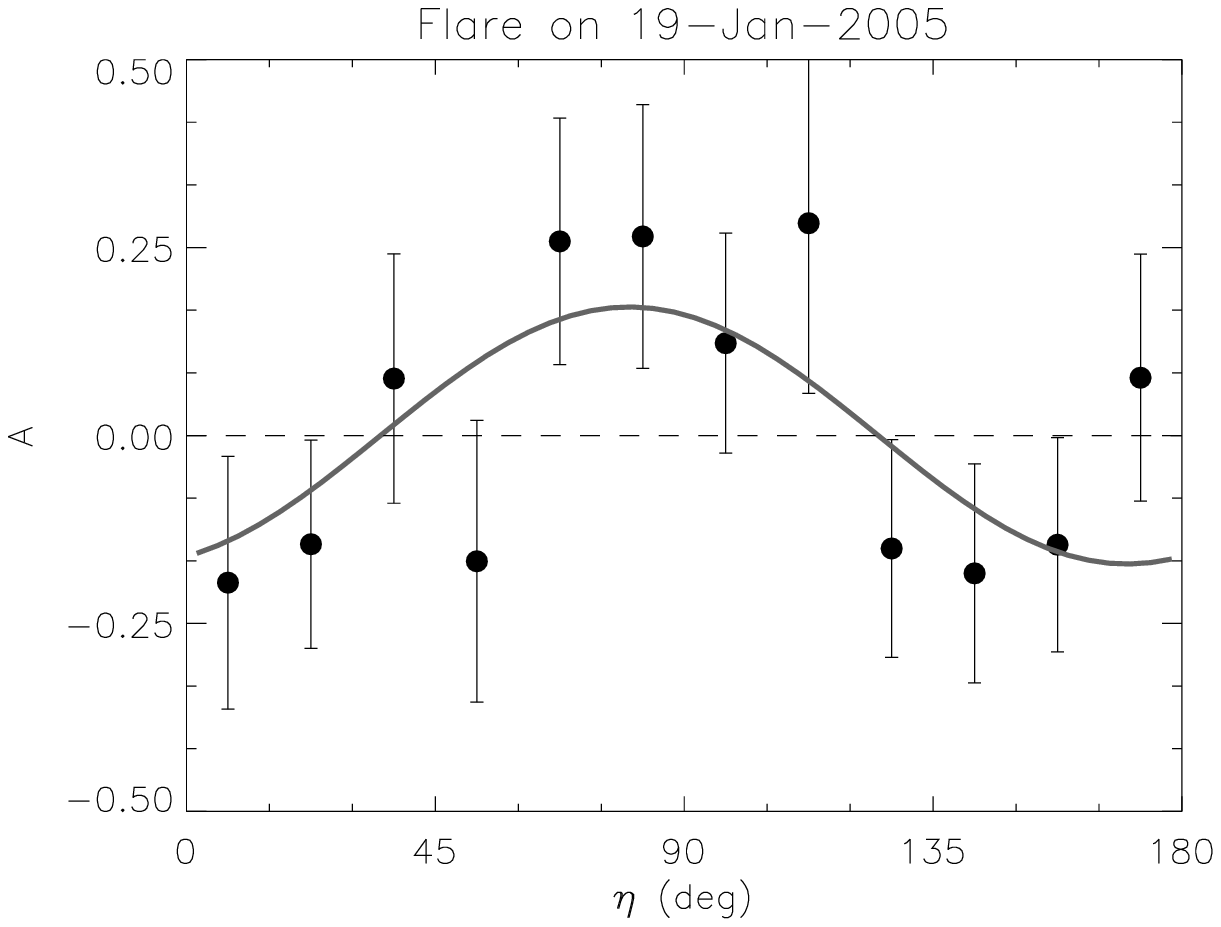,width=6.5cm}
	\epsfig{file=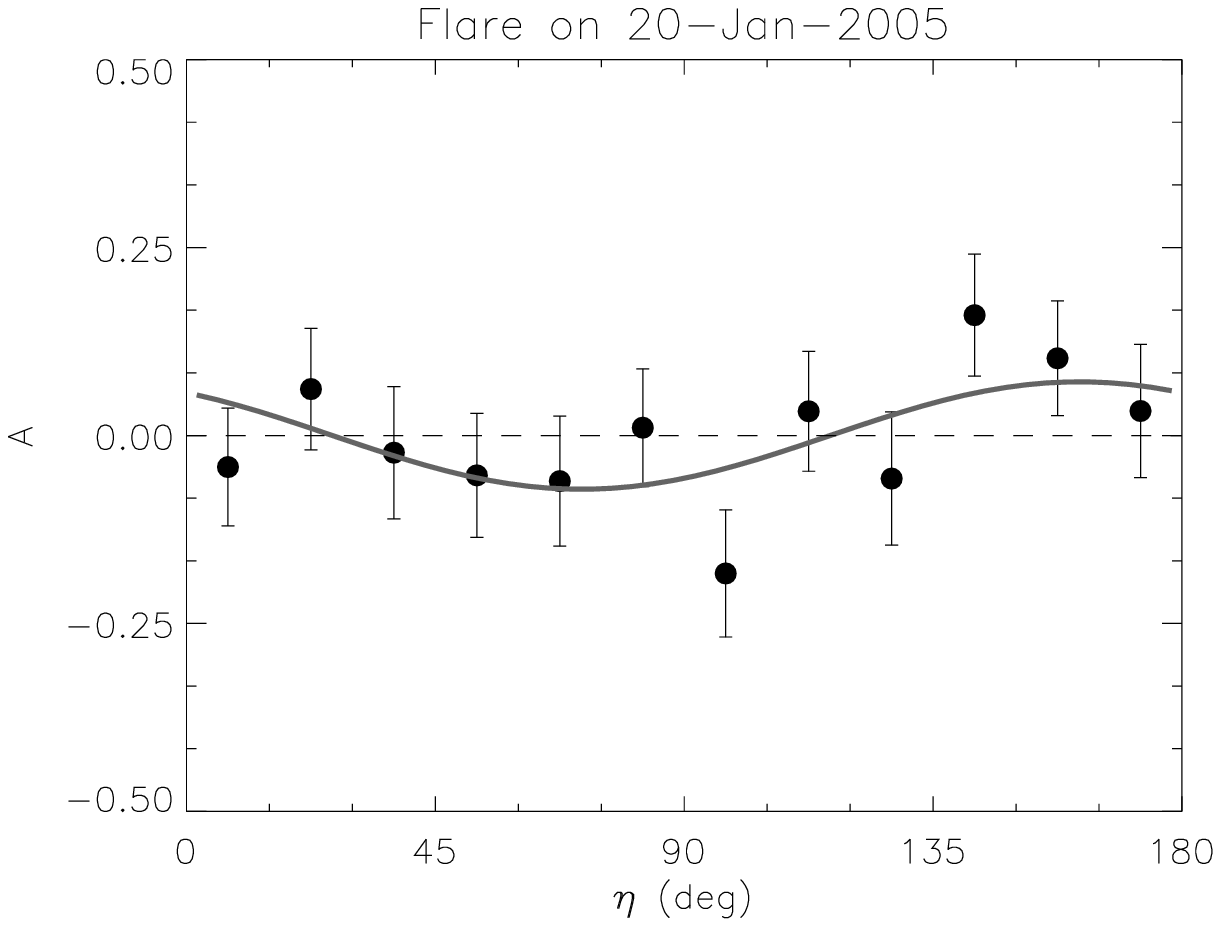,width=6.5cm}}
\hspace*{3.3cm}
\hbox{\epsfig{file=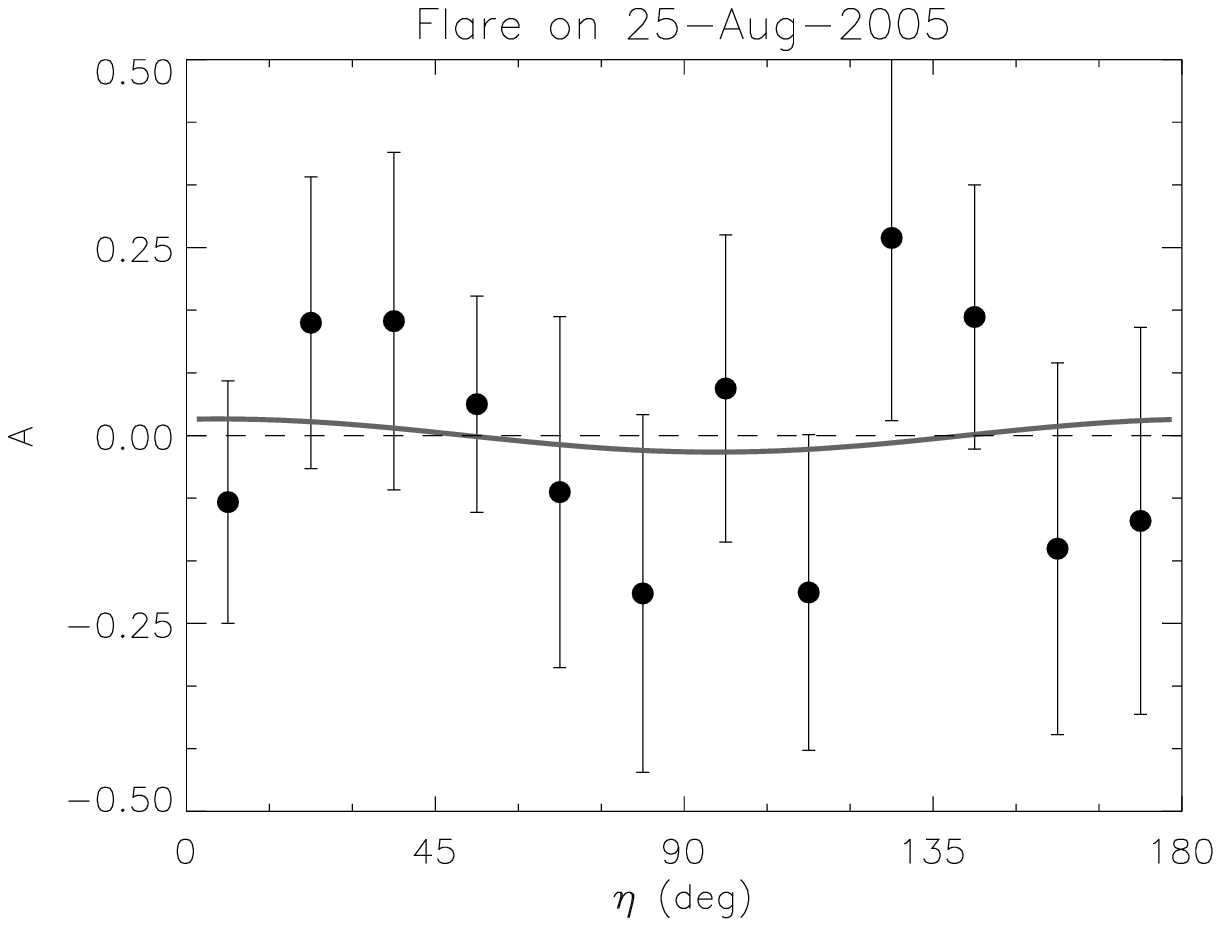,width=6.5cm}}
\caption{Asymmetry curves of all flares analyzed, extracted for photon energies between 100 keV
	and 350 keV. The \textit{thick lines} show the best fits with the function from Equation (\ref{eq:fitfun}). The
	\textit{angle of the minimum in the fit curve} indicates the flare polarization direction in heliocentric coordinates. }    
\label{fig:asy}
\end{figure}

The asymmetry curves $A(\eta_i)$ are displayed in Figure~\ref{fig:asy}, together with the best fit of the function in
Equation (\ref{eq:fitfun}). These curves have a periodicity of T/2 (T equal to RHESSI rotation period). To
improve the statistics, the second half of the asymmetry curves was added to their first 
half, plotting only the range $0^\circ$--$180^\circ$.

\begin{figure}			
\centerline{\epsfig{file=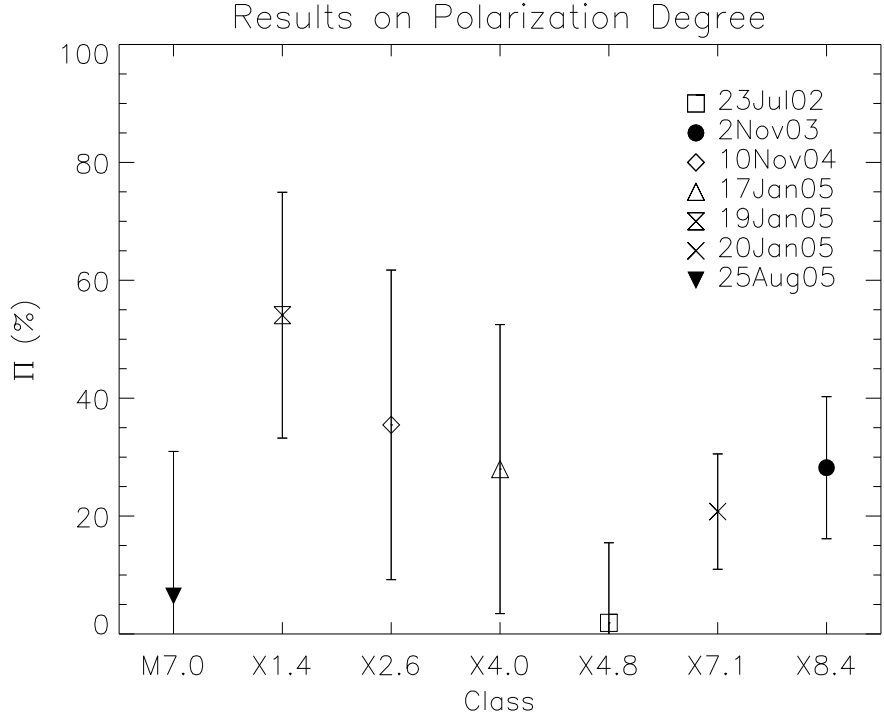,width=13cm,height=10cm}}
\caption{Results on the degree of polarization, with their $1\sigma$ error bars.}
\label{fig:results}
\end{figure}

The degrees of polarization $\Pi$ were calculated using Equation (\ref{eq:Poldeg}) with $\mu_{100}$
taken from simulations. The mean value of $\mu_{100}$ was 33.2\% and its variations between
different flares were below 1.1\%. The resulting polarization degrees were found to be between 2\% and 54\%, with
error bars varying from 10\% to 26\% at the $1\sigma$ level (see Table~\ref{tb:results}). The $\Pi$ values of all the
flares are plotted versus the flare class in Figure~\ref{fig:results}, where no significant correlation between these
two quantities can be observed.


\begin{landscape}	
\begin{table}[p] 		
\begin{tabular}{llllllll}
\hline
Flare number (RHESSI)	& 2072301	& 3110221	& 4111002	& 5011710	& 5011911	& 5012005	& 5082502	\\
Date			& 23 Jul. 2002	& 2 Nov. 2003	& 10 Nov. 2004	& 17 Jan. 2005	& 19 Jan. 2005	& 20 Jan. 2005	& 25 Aug. 2005 	\\\hline
$N_{tot}$		& $7439\pm86$	& $34723\pm186$	& $3816\pm62$	& $2142\pm46$	& $5688\pm75$	& $43313\pm208$	& $6139\pm78$	\\
$N_{acc}$		& $2269\pm10$	& $21427\pm31$	& $506\pm5$	& $473\pm5$	& $783\pm6$	& $26907\pm35$	& $602\pm5$	\\	
$N_{bg}$		& $1758\pm42$	& $5135\pm72$	& $2047\pm45$	& $733\pm27$	& $2784\pm53$	& $5808\pm77$	& $3717\pm61$	\\
$N_{C}$ 		& $3411\pm97$	& $8160\pm202$	& $1262\pm77$	& $937\pm54$	& $2121\pm92$	& $10598\pm225$	& $1820\pm100$	\\\hline
$\mu_{100}$ $(\%)$	& $33.0\pm1.6$	& $32.4\pm1.8$	& $32.9\pm1.7$	& $33.4\pm1.9$	& $31.6\pm1.5$	& $34.4\pm1.6$	& $34.4\pm1.6$  \\		
$\mu_p$ $(\%)$		& $0.6\pm4.5$ 	& $9.1\pm3.9$ 	& $11.7\pm8.6$	& $9.4\pm8.2$ 	& $17.1\pm6.5$	& $7.2\pm3.3$ 	& $2.2\pm8.4$	\\ 
$\phi$ (deg)		& $151\pm195$ 	& $96\pm12$	& $104\pm24$	& $71\pm29$	& $170\pm11$	& $66\pm14$ 	& $102\pm104$	\\
$\Pi$ $(\%)$		& $2\pm14$	& $28\pm12$	& $36\pm26$ 	& $28\pm25$	& $54\pm21$	& $21\pm10$	& $6\pm25$	\\\hline

\end{tabular}
\caption{Summary of polarization results for the flares studied. $\mu_p$ is the observed modulation factor. $\Pi$ is the 
	polarization degree of the flare, and $\phi$ its polarization angle given in heliocentric
	coordinates.} 
\label{tb:results}
\end{table} 
\end{landscape}


\begin{figure} 			
\centerline{\epsfig{file=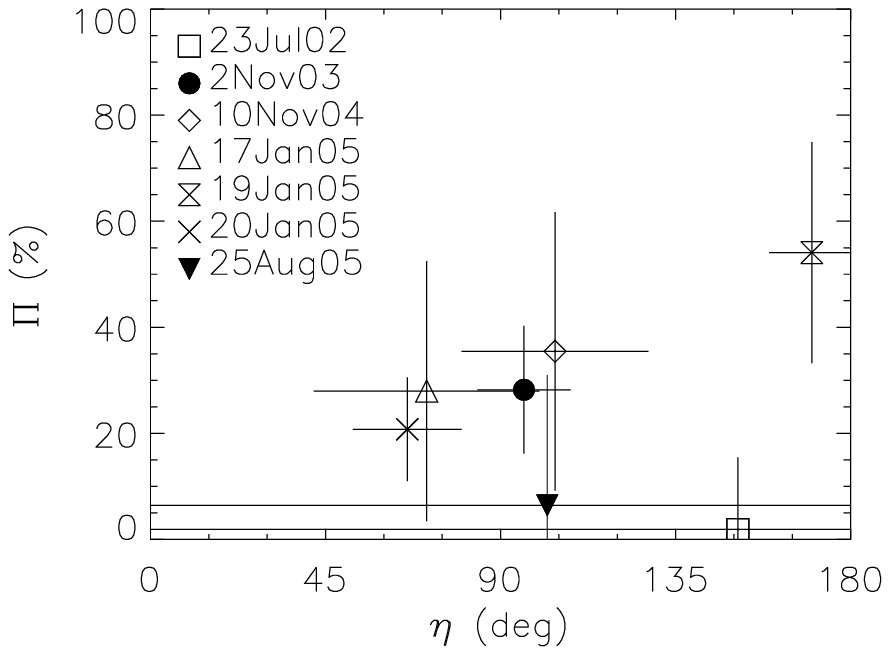,width=6.5cm}
	\epsfig{file=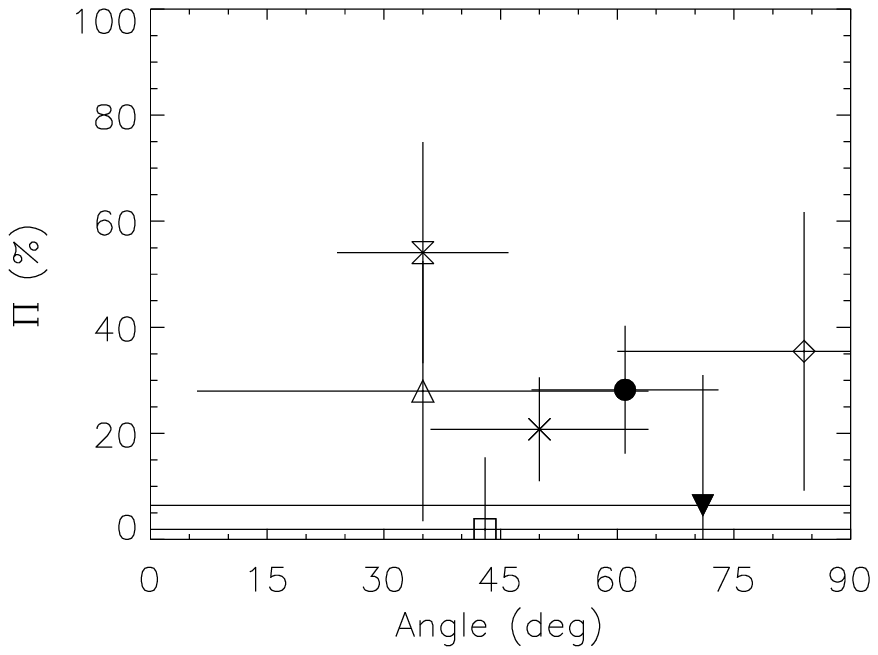,width=6.5cm}}
\caption{Polarization degree vs. polarization angle plotted in two different reference systems:
	with respect to solar equator (\textit{left}) and with respect to the radial line that joins the center of the Sun
	and the flare position (\textit{right}). } 
\label{fig:polvsang}
\end{figure}

The data does not show any preferential direction of polarization. When plotting the polarization angles of the flare
sample in heliocentric coordinates (Figure~\ref{fig:polvsang}, left), the points seem to concentrate around the
North-South solar direction but the error bars are too large to extract a firm conclusion. When the angle between the
polarization direction and the line that joins the flare position and the center of the Sun was calculated, no tendency
was found (Figure~\ref{fig:polvsang}, right).    

Despite the very high single-photon count rates of more than 20000 counts per second,
the mean number of Compton events found per flare was only around 4000, and the smallest one does not even reach
1000 coincidences. This reflects the small value of the RHESSI
effective area for Compton polarimetry as discussed in \S\ref{sec:method}. The signal-to-background ratio is on
the average only around 0.5. Depending on the flare, the largest background contribution is produced either by
accidental coincidences or by cosmic $\gamma$-ray background. The number of counts found with each flare
are compiled in the first part of Table~\ref{tb:results}, where the errors refer to 1$\sigma$ and are purely
statistical.

\subsection{Comparison with other measurements}
\label{subsec:compareMeasurements}

With respect to the polarization amplitude, our results are consistent with previous measurements made at higher
energies by \inlinecite{boggs06}. In the particular case of the 23 July 2002 flare that they also analyzed, our
value  ($2\%\pm14\%$) is smaller than theirs ($21\% \pm 10\%$), but agreement is found at the $1.5\sigma$ level. The
difference can be explained by the different time periods and energy ranges used in both cases. Extending the time 
and energy windows for our analysis towards  the values selected by \inlinecite{boggs06} provides very similar
polarization levels. The polarization angles of the two flares measured by \inlinecite{boggs06} are aligned along the
North-South direction in heliocentric coordinate system. However, these authors conclude that 
polarization is azimuthal for near-limb flares, but radial for those close to the Sun center. We can not confirm this
rule from our observations (see Figure~\ref{fig:polvsang}, right). The values found in the present work are more uniformly
distributed between $35^\circ$ and  $85^\circ$, independently of the flare location. Taking into account the values of the
error bars and the size of the statistical sample of analyzed flares, further measurements aimed to verify the observed
disparity are needed.    

Recent polarization data at energies up to 100 keV have became available from measurements with the
SPR-N instrument on board of the Coronas-F satellite \cite{zhitnik06}. From a sample of 25 solar flares,
upper limits on the polarization degree were found to be in the range from 8 to 40\% ($3\sigma$). These
values are in good agreement with our results (typically within $2\sigma$ level). In particular, for the
flare on 20 January 2005, observed simultaneously by both satellites, the polarization value from
RHESSI observations was equal to $21\%\pm10\%$ while the upper limit in the Coronas-F measurement was equal
to 17\%. Again, a more direct comparison is not possible because both the energy range and the time intervals analyzed
were different. For one flare (on 29 October 2003), the 
Coronas-F instrument showed a significant polarization degree that increases from about 50\% at energies
20--40 keV, up to more than 70\% for the energy channel 60--100 keV. Unfortunately, RHESSI polarization
analysis of the 29 October 2003 flare was not possible due to a high contamination of its detectors with
charged particles.

\section{Interpretation}
\label{sec:interpretation}

Our results were compared with theoretical predictions for the non-thermal photon emission given by
\inlinecite{bairamaty78}, \inlinecite{leachpetrosian83} and also with the 0\% polarization hypothesis.
\inlinecite{bairamaty78} provide the most comprehensive set of theoretical data, covering the whole energy
range from 10 to 500 keV. These authors considered primary X-ray emission due to Brems\-strah\-lung of the
accelerated electrons moving towards the photosphere, adding also the X-ray Compton backscattering
component. The photon polarization was studied for different electron spectra and pitch angle
distributions and results were presented as a function of the observing angle. At energies around
200 keV, typical for our analysis, the polarization reaches maximum values between 20 and 30\% and
decreases quickly for flares located closer to the solar center. Taking this trend into account, the predicted
polarization degrees for our sample of flares range from -19\% to 1\%, where the negative sign indicates a polarization
direction parallel to the magnetic field, and the positive one perpendicular to it. 

\inlinecite{leachpetrosian83} analyzed emissions from more complex, loop-shaped magnetic
fields, but most of their polarization prediction is given only for two energies: 16 and 102 keV. As the
dependence of the polarization on the energy was weak, we extrapolated their values
towards our energy range. The highest polarizations for large energies are expected to come either from
the top of the flare loop, or from the transition region above the chromosphere. The latter option roughly
corresponds to RHESSI observations in which the high-energy emissions come from the foot-points.
Nevertheless, a clear distinction between chromospheric and transition zone  emission was not possible.
\inlinecite{leachpetrosian83} proposed several models depending on the magnetic field gradient, the electron pitch
angle distribution and the spectral index. We selected three different cases for comparison purposes: one with
a homogeneous magnetic field and a uniform distribution of the electron pitch angles (model 3), another with equally
homogeneous magnetic field but pitch angles close to $90^\circ$ (model 5), and the last
one with a large magnetic field gradient and large pitch angles (model 8). After correcting for the
flare position on the Sun, the expected polarization degrees of our flares are in the ranges between
-7\% and 18\% for model 3, from -45\% to -80\% for model 5, and from -6\% to 23\% for model 8.  

The latest theoretical work on polarized emission, with the most advanced electron beam dynamics, was
presented by \inlinecite{zharkova95}. Polarization calculations were done only up to 100 keV (for photon
spectral indices typical to our flares), and showed increased values for higher energies. Their predictions, 
extrapolated to the energy range used in the present work, are similar to those from \inlinecite{bairamaty78} and
cannot be distinguished by the following $\chi^2$ analysis.

The comparison of our observations with the models was done by generating asymmetry curves for the flares polarized
in accordance with the theoretical predictions. Monte Carlo simulations were performed for each flare individually, 
using exactly the same RHESSI mass model as in \S\ref{sec:simulation}. The expected polarization angle and
amplitude were properly determined by adjusting the theoretical predictions to the flare
position on the Sun. The high number of simulated events allowed to keep the statistical errors on a very
low level comparing with the experimental data. Finally, the reduced $\chi^2$ value was calculated using
the data points from all seven flares as given by Equation (\ref{eq:chi2}). 

\begin{equation}  		
  \chi^2 = \frac{1}{n(m-1)} \sum_{i=1}^n\sum_{j=1}^m\frac{(N_{i,j}^e - N_{i,j}^t)^2}{(\Delta N_{i,j}^e)^2 +
  (\Delta N_{i,j}^t)^2},
\label{eq:chi2}
\end{equation}

\noindent where $i$ is the flare number and $j$ is the angular bin in the asymmetry plot. $N_{i,j}^e$ and $N_{i,j}^t$
are the experimental and theoretical numbers of coincidences respectively, and $\Delta N_{i,j}^e$ and $\Delta
N_{i,j}^t$ are their statistical uncertainties. The total number of degrees of freedom was equal to
77. The results for all four models and the 0\% polarization hypothesis are displayed in Table~\ref{tb:chi2}.

\begin{table}
\begin{tabular}{lc}
MODEL						& Reduced $\chi^2$ 	\\\hline	
\inlinecite{bairamaty78}			& 1.03 			\\
\inlinecite{leachpetrosian83} (model 3)		& 0.86 			\\
\inlinecite{leachpetrosian83} (model 5)		& 2.70 			\\
\inlinecite{leachpetrosian83} (model 8)		& 0.83 			\\
0\% polarization				& 0.82 			\\\end{tabular}
\caption{Reduced $\chi^2$ from all the flares combined, obtained by comparison of several theories with our
	measurements.}  
\label{tb:chi2}
\end{table} 
	
From the $\chi^2$ values of Table~\ref{tb:chi2} we can reject the model 5 from \inlinecite{leachpetrosian83}, with a
90\% of confidence. For the rest of the models the $\chi^2$ are very close to unity, preventing us from
distinguishing between them. Within the error bars, they all agree with the experimental data equally
well. The same is also valid for the 0\% polarization hypothesis. Further refinement would require a polarimeter able to
deliver data with error bars on the level of 1--2 percent.


\section{Summary and conclusions}
\label{sec:conclusions}

Measurements of hard X-ray polarization have been performed for six X-class and one
M-class flares from the RHESSI database. Our flare sample was identified after applying
selection criteria to the signal strength, background levels and level of contamination by charged particles
in RHESSI detectors. The selected energy range, from 100 to 350 keV, connects the
old (\opencite{tindo70}; \opencite{tindo72}; \opencite{tindo76}) and new \cite{zhitnik06} results at low energies with
the only measurement reported at a high energy band \cite{boggs06}.   

We found values for the polarization degree in the range between 2\% and 54\%, with statistical errors from
10\% to 26\% at the $1\sigma$ level. The polarization angles are distributed between $66^\circ$ and $170^\circ$ in
heliocentric coordinates. They do not show any preferential orientation of the polarization, neither parallel nor
perpendicular, with respect to the radial direction defined by the position of the flare in the Sun. In addition, no 
significant dependency between the orientation of the polarization and the distance of the flares to the Sun center was
found.  

The polarization orientation with respect to the line that joins the two major foot-points of the flare was also
studied. For this purpose, images of the flares were constructed with RHESSI at different energies. The emission
above $\sim$30 keV was found to be produced around the foot-points. However, no correlation between polarization
direction and foot-points orientation was found. Similarly, no relationship between solar flare intensity and
polarization degree could be observed.

The polarization degree from the 23 July 2002 flare measured by \inlinecite{boggs06} at high energies
is in agreement with our results at the $1.5\sigma$ level. Their conclusion about the orientation of
polarization respect to the radial direction passing trough the flare position (perpendicular to it for flares in
the limb and parallel for flares close to center) cannot be confirmed by our measurements. Our angles are distributed
between $35^\circ$ and  $85^\circ$ independently on the flare location. Comparison with theoretical predictions is more
complex, as the direction of the polarization is expected by the theory to change around 300 keV \cite{bairamaty78}.
Again, more observations with better accuracy are needed.   

Regarding the low energy measurements, our data are in good agreement with the recent results from the SPR-N
instruments on board of the Coronas-F satellite \cite{zhitnik06}, typically within $2\sigma$. Their large
sample of 25 solar flares reveals low polarization degrees, providing $3\sigma$ upper limits from 8 to 40\%.
Unfortunately, no information is given about the polarization direction. In the 29 October 2003 flare, for
which the Coronas-F team claims a polarization of 70\%, RHESSI measurements were contaminated by charged particles. 

Theoretical predictions of the non-thermal Bremsstrahlung emission provide polarization levels of the order of
20\%. However, depending on the assumptions used by different authors, the expected polarization can
differ not only by its value, but also by its orientation (see, for example, the calculations at 100 keV from
\inlinecite{bairamaty78} and \inlinecite{leachpetrosian83}). As the statistical uncertainties provided by our
instrument are in the same order as the model predictions, equally good agreement is found for any of them as well as
for the case of 0\% polarization. In order to distinguish between different models, polarimetry measurements of
at least 2\% accuracy are needed. Only model number 5 from \inlinecite{leachpetrosian83}, predicting polarization up
to 85\%,  could be rejected by our RHESSI data. In this model, the magnetic field has the same strength at
the top and the bottom of the loop, and the electrons spiral at pitch angles close to $90^\circ$. 

RHESSI has made the first steps towards the understanding of the polarization phenomena in solar flares above
100 keV, where non-thermal emission dominates. Measurements with accuracy better than 10 to 20\% were, however,
hardly possible. This is due to its small effective area and high levels of flare-induced 
background. Continuation of such studies will require a dedicated polarimeter that must solve the
problems inherent to the RHESSI design. Emphasis should be put on increasing the effective area and improving the
background rejection capabilities. In particular, a better time resolution will reduce the number of accidental
coincidences, and the optimization of the detector dimensions will improve the detection efficiency for the
Compton scattering. 


\end{article} 
\end{document}